# Mechanism of heteroepitaxial growth of boron carbide on the Si-face of 4H-SiC

Yamina Benamra[1‡], Laurent Auvray[1‡], Jérôme Andrieux[1‡], François Cauwet[1‡], Marina Gutierrez[2‡], Fernando Lloret[2‡], Daniel Araujo[2‡], Romain Bachelet[3‡], Bruno Canut[3‡], Gabriel Ferro[1‡]*

[1]Universite Claude Bernard Lyon 1, CNRS, LMI UMR 5615, Villeurbanne, F-69100.

[2]Materials Science Department, Universidad de Cádiz, 11510 Puerto Real, Spain.

[3]Institute of Nanotechnologies of Lyon, CNRS, UMR 5270, Villeurbanne, F-69100.

(*corresponding author: Gabriel FERRO; gabriel.ferro@univ-lyon1.fr)

‡These authors contributed equally.

KEYWORDS: boron carbide, heteroepitaxy, silicon carbide, CVD, semiconductors.

**ABSTRACT.**

Heteroepitaxial boron carbide ($B_xC$) can be grown on Si face 4H-SiC(0001) using a two-step process involving substrate boridation at 1200°C under $BCl_3$ + $H_2$ followed by a chemical vapor deposition (CVD) growth step at 1600°C by adding $C_3H_8$ precursor. However, in-depth investigation of the early growth stages revealed that complex reactions occur before starting the CVD at high temperature. Indeed, after boridation, the 35 nm $B_xC$ buffer layer is covered by an amorphous B-containing layer which evolves and reacts during the temperature ramp up between 1200 to 1600°C. Despite the formation of new phases (Si, $SiB_6$), which could be explained by significant solid-state diffusion of Si, C and B elements through the thin $B_xC$ layer, the CVD epitaxial re-growth upon reaching 1600°C does not seems to be affected by these phases. The resulting single crystalline $B_xC$ layers display the epitaxial relationships $[10\bar{1}0]B_xC(0001)\|[10\bar{1}0]4H\text{-}SiC(0001)$. The layers exhibit a $B_4C$ composition, e.g. the highest possible C content for the $B_xC$ solid solution.

## Introduction

Boron carbide is a technical ceramic which is mainly used in abrasive and refractory applications [1,2]. It is a solid solution, labeled $B_xC$ in this work, with a carbon composition ranging between ~9 at% and 20 at% [3]. It crystallizes in the trigonal system, space group $R\bar{3}m$, and its cell parameters a and c vary according to the C content: 560<a<564.5 pm and 1206<c<1218 pm [3]. The high thermal neutron capture cross-section of $^{10}B$ combined with the high stability (thermal, chemical and mechanical) of $B_xC$ have made it a promising candidate for neutron converter layers in solid-state neutron detectors [4]. On the other hand, despite being known as a semiconductor since decades [5,6], its basic electronic properties remain poorly known or even unexplored. For instance, its estimated bandgap, varying typically in the range of 1.6 to 2.2 eV, seems to depend on different parameters such as its chemical composition or the presence of impurities [7,8]. The difficulty of growing high-quality crystals, either in bulk or thin film forms, is another major challenge to solve before considering $B_xC$ for any electronic application. This challenge stems from $B_xC$'s complex crystal structure (combination of icosahedral $B_{12}$ or $B_{11}C$ units and linear C-B-C or C-B-B chains, depending on atomic composition [9,10]) and from the required high growth temperatures ($B_xC$ melting point ~2763 K [11]). Several traditional methods for synthesizing $B_xC$ include carbothermal reduction, magnesiothermic reduction, direct synthesis from pure elements, and synthesis from polymeric precursors [12]. These techniques typically produce polycrystalline and often impure material. For single-crystal growth, the primary methods are the floating zone technique [13-15] and growth from metal solutions [3,16]. However, both methods face challenges such as significant contaminants incorporation (solvent, oxygen) in the grown crystals or limitations in crystal size.

$B_xC$ thin films can be deposited using physical vapor deposition (PVD) techniques [17-19], typically resulting in amorphous layers due to low deposition temperatures (<1000°C). The chemical vapor deposition (CVD) of $B_xC$ has also been extensively studied at higher temperatures (1000-1600°C) using mainly boron halide ($BCl_3$) and methane ($CH_4$) or tetrachloromethane ($CCl_4$) [20-23]. Depending on the experimental conditions used, the $B_xC$ microstructure ranged from amorphous to polycrystalline, with significant variations in stoichiometry in relation with precursors ratio used in the gas phase [24,25]. Note that $B_xC$ heteroepitaxy was not targeted in these studies which used poorly adapted substrates (graphite, silicon, tungsten…).

When focusing on the heteroepitaxial growth of $B_xC$, 4H-SiC substrate seems a good choice due to the chemical, thermal, and crystallographic compatibilities (hexagonal symmetry) between the two materials, despite the huge lattice mismatch (82%). Recent research has shown that epitaxial $B_xC$ films can indeed be grown by CVD on the C-face $(000\bar{1})$ of 4H-SiC using single-source precursor (triethylborane - TEB) at temperatures between 1300°C and 1500°C [26,27]. However, co-deposition of graphite occurred at these high temperatures, degrading the purity and quality of the $B_xC$ layers. On the other hand, attempts to grow $B_xC$ on the Si-face of 4H-SiC crystals have been less successful, almost systematically resulting in polycrystalline or amorphous films [26-28], the reason being still unclear. But, as 4H-SiC electronics is almost exclusively fabricated on the Si face, one would clearly need to succeed in B$_x$C heteroepitaxy on this surface polarity in order to take full advantage of using this substrate. Toward this end, we have recently demonstrated that heteroepitaxy of B$_x$C on Si-face 4H-SiC can be obtained using a two-step CVD process [29] which involves the insertion of a boridation step before CVD regrowth. These preliminary results, which were quite descriptive, have shown that the growth mechanism allowing such heteropitaxy is rather complex and that further work would be required for reaching fine understanding of it. This is the main goal of the present study which explores in more details the role and evolution of the B$_x$C buffer layer formed during the boridation step, prior to conventional CVD growth.

**Experimental setup**

B$_x$C films were grown in a homemade vertical cold-wall CVD reactor operating at atmospheric pressure. The SiC-coated graphite susceptor, on which the samples were placed, was RF-heated and the temperature was regulated using an optical pyrometer in the range 600°C - 1600°C. Boron trichloride (1% BCl$_3$ in Ar, 99.999% purity) and propane (5% C$_3$H$_8$ in H$_2$, >99.95% purity), boron and carbon precursors respectively, were diluted in high-purity hydrogen carrier gas. We chose different precursors for C and B elements (unlike single-source TEB used in refs [26-27]) in order not only to possibly tune the C/B ratio in the gas phase during CVD but also to allow using B precursor separately for the boridation step, as described below. The depositions were performed on commercially available Si-face 4H-SiC(0001) substrates with the standard 4° off-orientation towards [11$\bar{2}$0]. Prior to growth, substrates were ultrasonically cleaned 10 min in methanol before being introduced inside the reactor via a loading chamber purged with argon. Before each deposition, the 4H-SiC crystals were treated under $H_2$ at 1000°C for 5 min to remove the native oxide. Then the rest of the growth procedure could start, as described below.

The goal of the boridation step is to form a thin heteroepitaxial $B_xC$ layer on top of the 4H-SiC substrate, similarly to the carbonization step commonly performed when growing 3C-SiC on silicon substrate [30]. It consists in heating the 4H-SiC substrate at 1200°C for 10 minutes under 2.5 sccm of $BCl_3$ only (no propane). On this $B_xC$ buffer layer, epitaxial re-growth can be done by adding propane to the gas phase. This CVD step was carried out at 1600°C for 60 minutes with a C/B ratio in the gas phase equal to 1 ($BCl_3$ = 2.5 sccm, $C_3H_8$ = 0.83 sccm). The transition between boridation and CVD was performed under $H_2$ with a 10°C/s ramp up rate. For comparison purpose, $B_xC$ layers were also grown directly at 1600°C without the boridation step, using the same precursor flows (introduced upon reaching the targeted temperature).

Samples were routinely characterized using optical Nomarski microscopy (Olympus BX60), scanning electron microscopy (SEM - FEI Quanta 250 FEG), X-ray diffraction (XRD - Bruker D2 PHASER Benchtop, θ-θ Bragg-Brentano configuration, Kα, λ=0.154 nm), and atomic force microscopy (AFM - Nano-Observer® microscope) in tapping mode. Energy dispersive Spectroscopy (EDS) were performed during SEM imaging using 2 keV acceleration beam, which allows probing around 80 nm of material. Growth rates were calculated by cross section measurement of the deposits thickness with SEM. Cross-sectional high-resolution transmission electron microscopy (HR-TEM) with a FEI TALOS F200X microscope operating at 200 kV was employed to investigate the epitaxial relationship between selected $B_xC$ layers and the 4H-SiC substrate. Detailed high resolution complementary XRD characterizations (ω-scan, φ-scan and pole figure) were performed using a five-circles Smartlab© diffractometer (Rigaku) equipped with a high-brilliance (9 kW) copper rotating anode source and two-bounces Ge(220) monochromator (K$\alpha_1$, $\lambda$=0.1540598 nm). A 473 nm laser-diode, focused to form a spot of few µm² section area, was used for micro-Raman spectroscopy measurements. Finally, Rutherford Backscattering Spectrometry (RBS) was performed at the SAFIR platform located at Paris Institute of Nanosciences. The backscattered ions were detected with a 13 keV resolution implanted junction set at an angle of 165° with respect to the beam axis. In order to probe the entire thickness of the $B_xC$ layer, $H^+$ ions of 1.5 MeV energy were used. The raw data of the analysis was processed using SIMNRA [31] simulation code in order to obtain the average composition of the layer and its thickness (assuming a theoretical density of 2.52 g.cm$^{-3}$ for $B_xC$ material).

## Results and discussion

As reported earlier, direct CVD growth at 1600°C on Si-face 4H-SiC substrate and using the same precursors system leads systematically to polycrystalline $B_xC$ deposit [29]. When inserting a boridation step before CVD at 1600°C, the deposit is homogeneous and completely covering with serrated steps morphology (triangles pointing in the same direction, Fig. 1a). This specific morphology is better seen by AFM (Fig. 1b). The root-mean-square (rms) roughness is rather high, close to 30 nm for 50×50 µm² area and 8.8 nm for 5×5 µm² area, as measured by AFM. Macroscopically, it gives a milky appearance to the layer by naked eye.

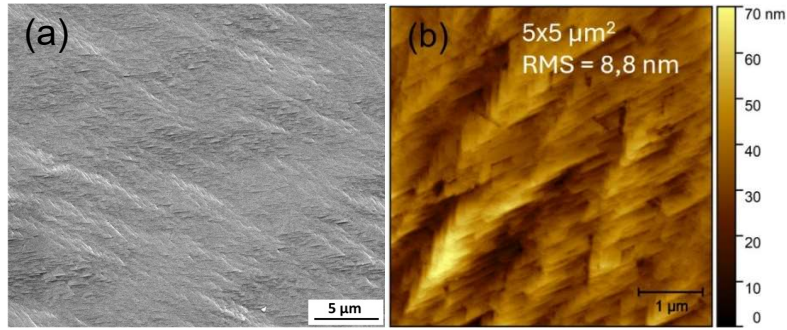

**Figure 1.** a) SEM images of a $B_xC$ film obtained on 4°off 4H-SiC after the two-step growth procedure; b) AFM image (5×5 μm² scan) of the same sample.

The diffractogram obtained on this sample, after aligning with respect to the symmetric planes of the substrates (maximizing the substrate signal), shows sharp and intense $B_xC$ peaks exclusively associated with the (000l) planes (Fig. 2a). The absence of other $B_xC$ reflections than the (0001) ones indicates a preferential out-of-plane orientation of the $B_xC$ layer along the c-axis [0001]. The crystalline mosaicity of this $B_xC$ layer, estimated from the full width at half maximum (FWHM) of the ω-scan around the $B_xC$(0003) plane, is relatively low (around 0.5° for the rocking curve, inset in Fig. 2a) which strongly suggests an epitaxial relationship between $B_xC$ and 4H-SiC.

In-plane XRD crystallographic analysis (φ-scan) was challenging due to the 4°off misorientation of the substrate. To address this, the same two-step growth process was repeated on an on-axis 4H-SiC(0001) Si-face substrate. This on-axis deposit displayed a similar, yet rougher morphology (Fig. 3) compared to the off-axis counterpart. XRD analyses in φ-scan geometry were performed on this sample using $(10\bar{1}1)$ and $(10\bar{1}4)$ planes for the substrate and the $B_xC$ layer respectively. An in-plane orientation relationship between the two materials was demonstrated (Fig. 2b). Moreover, despite the intrinsic three-fold symmetry of the rhombohedral $B_xC$ structure ($R\bar{3}m$ space group), an actual six-fold symmetry was evidenced. This indicates that the deposit is twinned with two types of domains (A and B) exhibiting 120° angular periodicity. The difference in reflection intensities suggests that A-type domains are dominant. In-plane mosaicity, so-called twist, of the domains A and B revealed by the φ-scan around the $(10\bar{1}4)$ reflection (FWHM values) are significantly higher (1.3° and 1.4°, inset in Fig. 2b) than the out-of-plane mosaicity (0.5° from ω-scan. These higher values support the hypothesis of twinning defects within the $B_xC$ layer grown on on-axis 4H-SiC(0001) Si face.

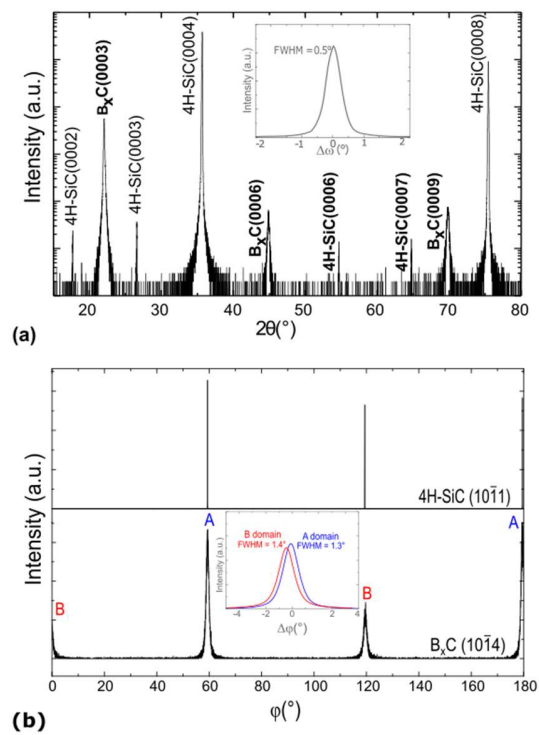

**Figure 2.** a) X-ray diffractogram of the $B_xC$ film grown on 4°off 4H-SiC using the two-step procedure. The inset shows the rocking curve (ω-scan) for the (0003) reflection of the same film. b) φ-scans of $B_xC$ grown on on-axis 4H-SiC using the same procedure, for $(10\bar{1}1)$ 4H-SiC substrate reflections and $(10\bar{1}4)$ $B_xC$ reflexions. The inset shows the fitted rocking curve in φ for the $B_xC(10\bar{1}4)$ reflections of the two twinned domains A and B.

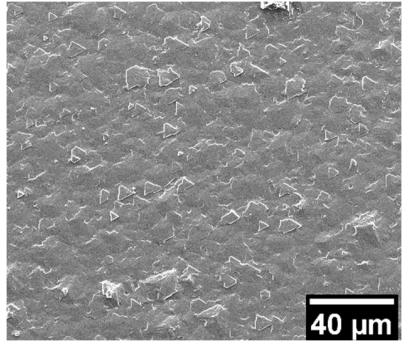

**Figure 3**. SEM image showing the morphology of the B$_x$C layer grown on on-axis 4H-SiC with the two-step growth procedure.

Pole figure (Fig. 4) confirms the well-defined B$_x$C layer epitaxy on the on-axis substrate. Characteristic (10$\bar{1}$4) B$_x$C reflections appear at φ = 0°, 60°, 120° and 180°, and at χ ≈ 58°, corresponding to the angle between (10$\bar{1}$4) and (0001) planes. No other parasitic preferential orientation is observed. As a conclusion, XRD analyses demonstrate the following single epitaxial relationship: B$_x$C(0001)[10$\bar{1}$0] ∥ 4H-SiC(0001)[10$\bar{1}$0], which confirms previous results reported for heteroepitaxy on the C-face of 4H-SiC [26].

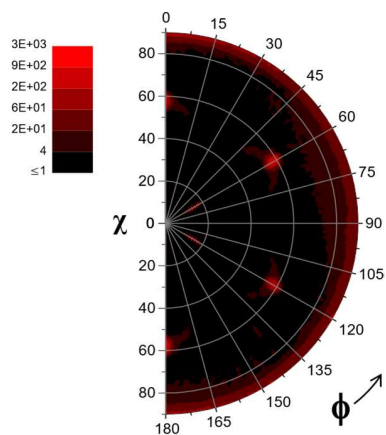

**Figure 4.** Pole figure of the (10$\bar{1}$4) planes family of the B$_x$C epilayer grown on on-axis 4H-SiC.

Figure 5 shows a high-magnification cross section TEM image of the 4° off-axis heteroepitaxial layer (~1.8 μm thick). The interface is not atomically smooth but it does not contain any features such as foreign phase or polycrystalline inclusion. The layer exhibits a high density of extended crystal defects which decreases along the film thickness [29]. The selected area electron diffraction (SAED) patterns recorded for both the substrate and deposit confirm the preferential film orientation: $B_xC(0001) \parallel 4H\text{-}SiC(0001)$. These results suggest that the CVD regrowth after the boridation step occurred smoothly.

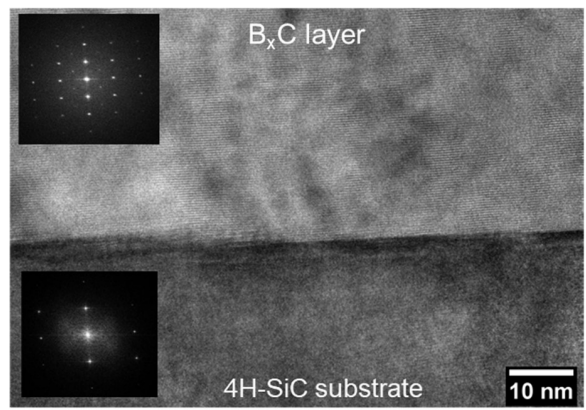

**Figure 5.** Cross sectional TEM image of the $B_xC$ deposit on 4H-SiC 4° off-axis Si-face, taken at the substrate/epilayer interface. Inserts correspond to the corresponding SAED of the probed area (upper for the $B_xC$ epitaxial layer and below for the substrate).

The areal mass and the mean elemental composition of a typical heteroepitaxial $B_xC$ layer were extracted from RBS analysis. A typical example is displayed on figure 6. Starting from the raw spectrum, the deconvolution of the C and B elemental contributions suggests a C content [C] = 20 ± 1 at% in the layer. One can thus consider that the heteroepitaxial layer has the $B_4C$ stoichiometry. Note that some traces of Si were also detected inside the $B_xC$ layer, but with a content in the 0.3-0.5 at% range (close to the detection limit of the technique). In this example, the areal mass of the deposit was found to be 250 µg.cm$^{-2}$, which corresponds to an equivalent thickness of 1.0 µm of bulk B$_4$C in correct agreement with the expected thickness from the elaboration procedure.

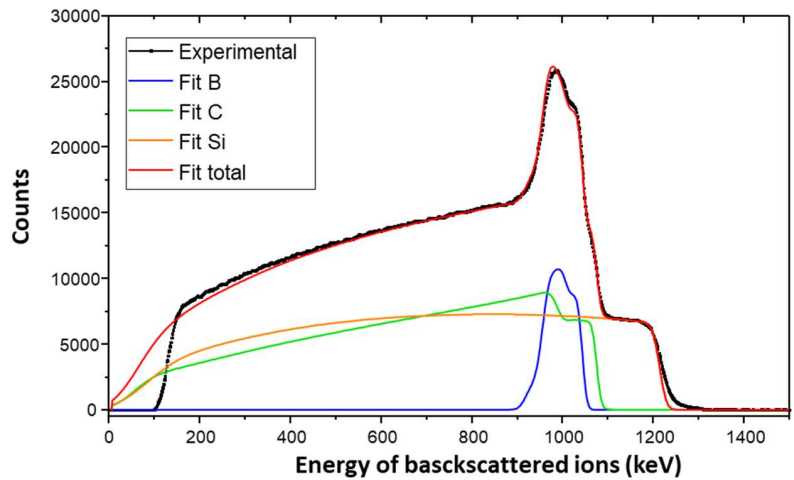

**Figure 6.** RBS spectrum and simulated data of B$_x$C heteroepitaxial layer grown on 4H-SiC, with the C, B and Si elemental contributions. Analysis conditions: H$^+$ ions of 1.5 MeV energy; detection angle: 165°. The solid curves are fittings using the SIMNRA simulation code [31].

*Understanding of the growth mechanism*

As confirmed by the above results, the boridation step is crucial for heteroepitaxial growth of B$_x$C on 4H-SiC. Toward a better understanding of this role, a growth experiment was stopped just after the boridation step. The morphology of such sample is very smooth and homogeneous as observed by SEM (Fig. 7a). AFM image confirms the low roughness of the surface after boridation (~1.2 nm RMS for a 50 × 50 µm² scan and 0.9 nm for 1 × 1 µm², Fig. 7b). The morphology is composed of small and agglomerated nodules of approximately 10 nm in size. Cross sectional TEM characterization of this sample (Fig. 8a) revealed the presence of two distinct layers: a thin (35 nm) and slightly rough one in direct contact with the SiC substrate and a thicker one (155 nm) above it, appearing smooth and featureless. The associated SAED patterns (Fig. 8b) show that the interface layer is composed of single-crystalline ((0001) oriented) $B_xC$ though with high defects density and twinning. The layer above it was found amorphous to polycrystalline (see diffuse rings in SAED pattern). EDS analyses indicate that the amorphous layer covering the $B_xC$ interfacial layer is mainly composed of boron while containing also traces (few at%) of silicon and carbon (Fig. 8c-d). The presence of oxygen can be reasonably attributed to contamination upon air exposure of the thin lamella before TEM analyses.

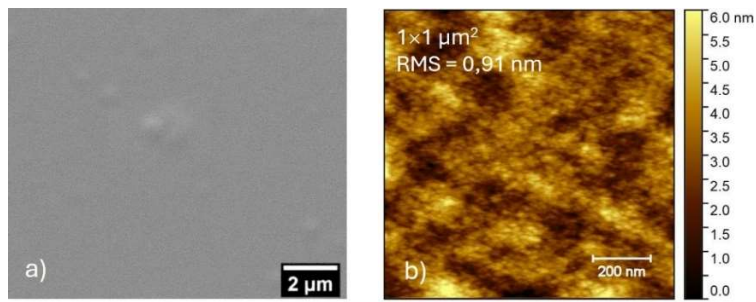

**Figure 7.** a) SEM and b) 1×1 µm$^2$ AFM image of sample surface after boridation.

From these results we can thus speculate that the boridation works similarly to the carbonization step in the 3C-SiC/Si heteroepitaxy, by forming a heteroepitaxial carbide seed at the substrate surface. Of course, the presence of an amorphous top layer containing mainly B (for the boridation case) makes the two processes somehow different. $B_xC$ formation at the substrate interface should result from the direct reaction of $BCl_3$ with SiC, which serves as the sole carbon source at this stage of the process (no propane). One can propose the following (unbalanced) chemical reaction describing $B_xC$ formation:

$$H_{2\,(g)} + BCl_{3\,(g)} + SiC_{(s)} \rightarrow B_xC(Si)_{(s)} + Si_{(s)} + HCl_{(g)}$$

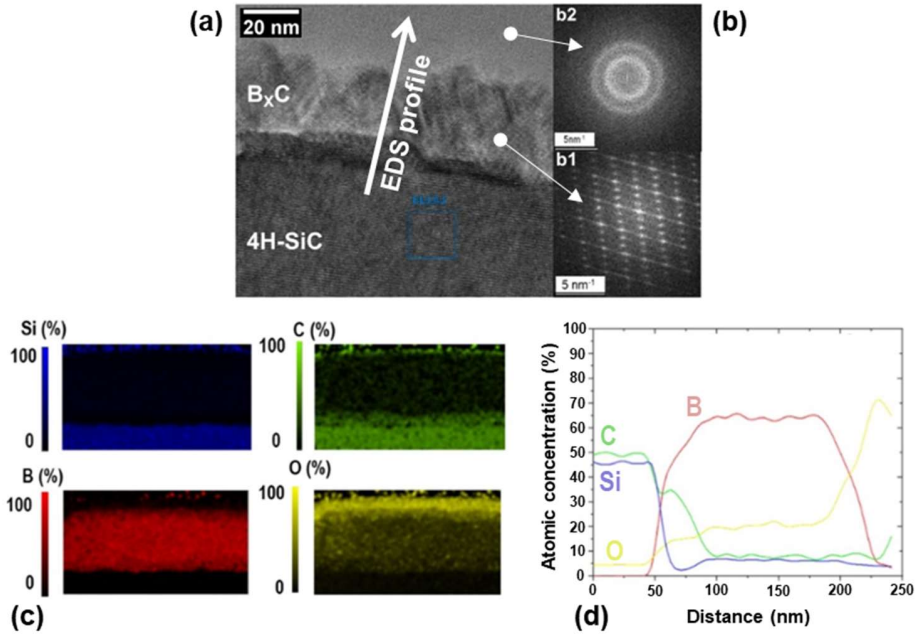

**Figure 8.** TEM cross-sectional image of the sample after boridation: a) HRTEM image centred on the $B_xC$ layer, b) associated SAED patterns of the $B_xC$ layer (b1) and the upper layer (b2), c) EDS elemental maps of Si, B, C, and O recorded on the cross-section of the sample, and d) EDS profiles of atomic compositions across the stack.

Despite being unbalanced, this equation suggests that $B_xC$ formation should lead to Si release inside the system. TEM analyses did not allow detecting any Si-containing inclusions except some traces of this element inside the amorphous layer (Fig. 8c and d). We can speculate that most of the released Si atoms can be eliminated under the form of gaseous species by the reaction with either H or Cl (or both) elements. The formation of B excess on top of the initially formed $B_xC$ layer could be explained by a slowing down of the above reaction which could become diffusion limited (at 1200°C) through the initially formed $B_xC$ layer. This slow diffusion could concern either C and Si (out-diffusion) or B (in-diffusion) atoms. This should lead to amorphous B accumulation at the surface as a result of BCl$_3$ cracking. Nevertheless, this amorphous layer does not prevent further B$_x$C epitaxy from occurring during the CVD step.

To investigate the changes occurring during the transition from boridation to CVD, a sample was prepared using the two-step procedure but stopping the experiment just before the introduction of precursors upon reaching 1600 °C, i.e. after the temperature ramp up from 1200 to 1600 °C under H$_2$ at 10 °C/s. The surface of the resulting sample is composed of three different phases, which could be better distinguished using the Z-contrast in backscattered electron (BSE) SEM imaging (Fig. 9a): i) light-grey big facetted crystals (5-15 µm lateral size), ii) white small nodules (~0.5 µm size) and iii) a dark grey phase obviously covering completely the 4H-SiC surface. Si being the heaviest element of the present chemical system, the brighter the grey contrast, the higher the Si content of the phase.

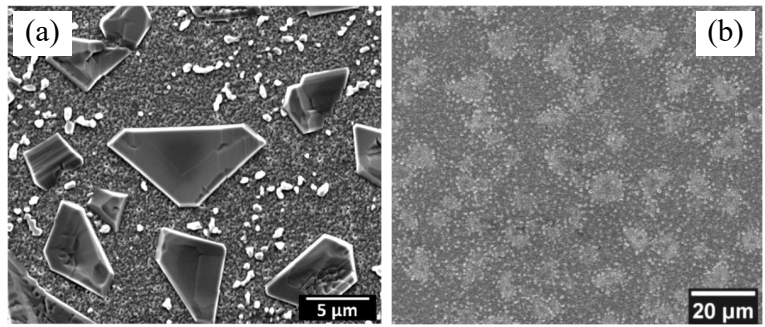

**Figure 9.** SEM images with BSE contrast of a) a sample prepared using the two-step procedure but stopped just before the CVD and b) the same sample after HF/HNO$_3$ etching.

The white nodules in figure 9a were indeed identified as pure Si by µ-Raman spectroscopy (Fig. 10a). XRD analyses allowed also detecting this pure Si phase together with epitaxial B$_x$C (Fig.

10b). Surprisingly, no additional peak coming from the light grey facetted crystals was detected by XRD. Chemical etching of the sample surface in concentrated HF/HNO$_3$ solution at room temperature led to disappearance of both the faceted crystals and the Si nodules, leaving only imprints of their previous presence (Fig. 9b). Since it is known that SiC and B$_x$C are chemically resistant to such etching, this result indicates that the facetted crystals are not composed of one of the two latter compounds. They could potentially be borides or silicides (or solid solutions out of these elements). EDS analysis allowed clarifying this point with the qualitative detection of mainly Si and B elements (Fig. 11). A rough attempt to quantification without physical standards led to an estimation of the chemical composition of the faceted crystals close to that of SiB$_6$ phase.

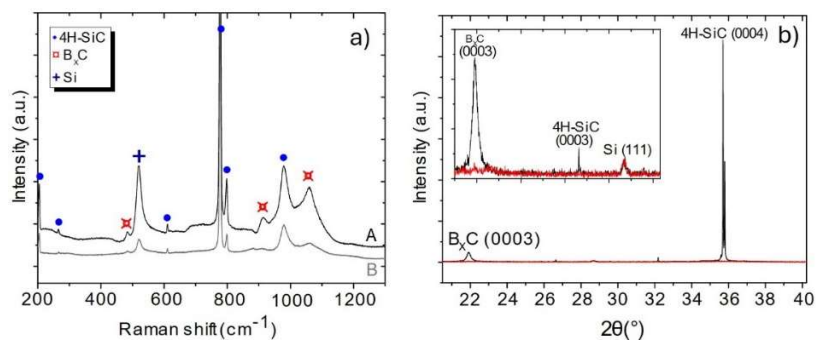

**Figure 10.** a) µ-Raman spectra of the sample shown in Figure 9a recorded at two different areas: (A) between the facetted crystals and (B) on the faceted crystals. b) X-ray diffractograms recorded on the same sample, with substrate signal maximized (black curve) and minimized (red curve); the inset shows a zoomed view of the diffractogram between 2θ = 21° and 30°.

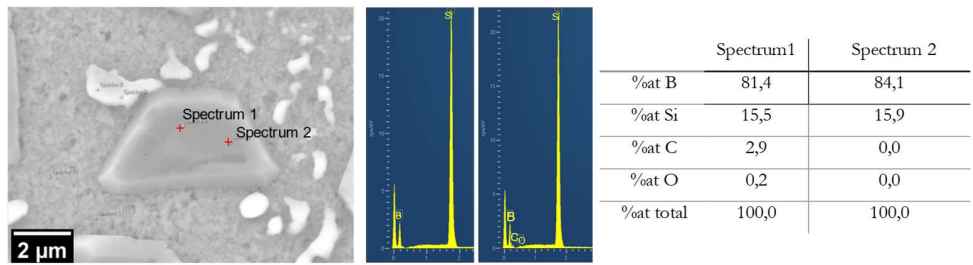

**Figure 11.** EDS analyses (under 2 keV electron beam) performed on the same sample as in Fig.9. The location of the focus points is presented, along with the corresponding EDS spectra and approximate quantification results without physical standards. The electron penetration depth was estimated to be less than 80 nm only, so that no signal comes from underlying phases.

TEM cross-sectional analysis (Figure 12) on the same sample showed that the interfacial $B_xC$ layer it still epitaxial but it has now thickened up to ~150 nm. It is covered by a phase (most probably a faceted $SiB_6$ crystal) with low diffraction contrast, having a mixed crystalline/amorphous nature. The thickening of the $B_xC$ layer without adding propane to the gas phase suggests that the additional C atoms required for this growth are brought by the substrate, probably by out-diffusion through the existing $B_xC$ layer during the temperature ramp from 1200 to 1600°C. Since no Si accumulation is detected at the $B_xC$/SiC interface, the Si atoms released by this C out-diffusion should also out-diffuse to the $B_xC$ surface. This was experimentally demonstrated by the detection of Si containing phases ($SiB_6$, Si) at the surface. Of course, there is no reason to exclude any B in-diffusion from the surface to $B_xC$/SiC interface, which should also participate to $B_xC$ layer growth.

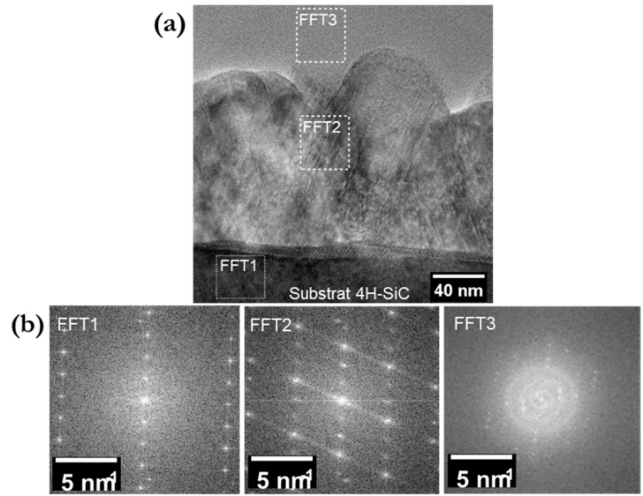

**Figure 12.** Cross sectional TEM characterizations of a sample prepared using the two-step procedure but stopped just before the CVD: a) high-resolution image of the $B_xC/SiC$ interface and b) SAED patterns corresponding to the three probed regions in a): the substrate (FFT1), the $B_xC$ layer (FFT2), and the phase above $B_xC$ (FFT3).

The last important question to tackle is the absence of obvious detrimental effect of the Si nodules and the facetted $SiB_6$ crystals on further $B_xC$ epitaxial growth by CVD. Everything seems to happen as if these Si-containing phases were not present at the surface upon addition of $BCl_3$ and $C_3H_8$ at 1600°C to start the CVD growth. We speculate that the excess of Si and B at the surface could be etched away, probably by the chlorinated by-products generated by $BCl_3$ cracking. Indeed, in the case of Si, etching of this element by chlorinated species at high temperature is a commonly known reaction for SiC surface preparation [32]. Note that, since the transition from 1200°C to 1600°C was performed under $H_2$ only, the presence of Si excess upon reaching 1600°C indicates that $H_2$ only is not enough for removing this Si excess, which confirms the need of Cl-containing species for etching Si atoms. In the case of $SiB_6$, the B atoms could serve as additional B source for $B_xC$ growth under propane flux, while Si atoms could be etched away by Cl-containing by-products. In overall, these reactions could be so fast at 1600°C that it would not affect the surface diffusion and incorporation of incoming B and C atoms at epitaxial sites. All the processes and hypotheses discussed above are summarized in Figure 13.

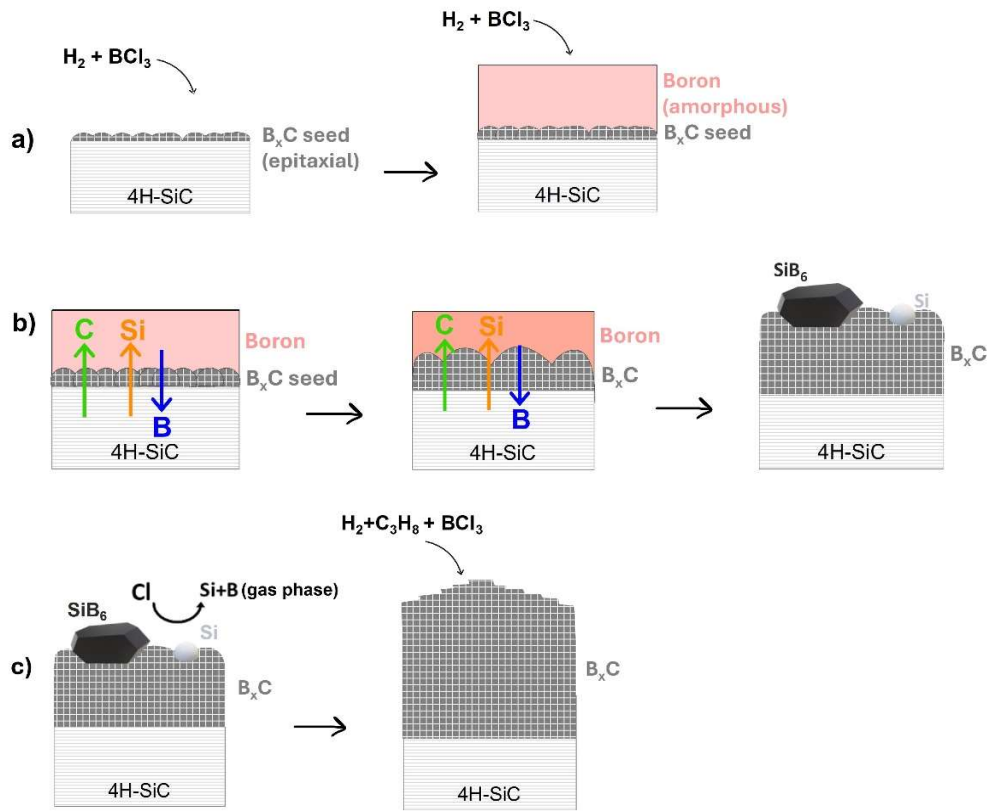

**Figure 13.** Mechanisms proposed for the various steps of the growth procedure. a) Boridation at 1200°C, b) Ramp-up transition (1200 to 1600°C) under $H_2$ and c) CVD at 1600°C.

In conclusion, though the boridation step is not as simple as the carbonization for 3C-SiC on Si case, it has a crucial role by nucleating the initial $B_xC$ in epitaxial relationship with the 4H-SiC (Si-face) substrate. Such heteroepitaxial nucleation on Si face cannot be done using direct CVD. Surprisingly, the main side effect of this boridation (amorphous B layer on top of $B_xC$) participates positively to the epitaxial growth by providing B atoms for further thickening of the initial $B_xC$ layer (by cross diffusions) during the temperature ramp up. The excess of foreign elements (B and Si) at the surface does not negatively impact the CVD re-growth on top of the already formed $B_xC$ layer. The processes leading to their elimination are still unclear, but they are efficient enough to occur at the early stage of CVD regrowth without leaving any trace of their transient presence. Finally, despite the complexity of the various processes occurring, the two-step approach developed in the present study is robust and efficient enough to reproducibly grow heteroepitaxial $B_xC$ layers on the Si face of 4H-SiC substrate.

## Conclusion

In this study, we have investigated in details the mechanism involved in the boridation step allowing heteroepitaxial growth of boron carbide by CVD on Si-face 4H-SiC(0001). The ramp up step after boridation is accompanied by several transient processes (chemical reactions and solid-state diffusion) which further consumes the B excess on top of the buffer layer and allow starting the CVD growth in apparently unaffected conditions. The epitaxial relationship of $[10\bar{1}0]B_xC(0001)\|[10\bar{1}0]$4H-SiC(0001) has been confirmed. The layers exhibit a $B_4C$ stoichiometry. This work opens the way to practical uses of crystalline $B_xC$ material for semiconductor applications, either as a layer or in heterostructure with 4H-SiC.

## Acknowledgment

Electron microscopy was performed at the Centre Technologique des Microstructures de l'Université Claude Bernard Lyon1 (CTµ).

**Author Contributions**

The manuscript was written through contributions of all authors. All authors have given approval to the final version of the manuscript.